\documentclass[12pt,preprint]{aastex}

\begin{document}

\title{Direct measurement of the size of 2003 UB313 from the Hubble Space Telescope}
\author{M.E. Brown\altaffilmark{1}, E.L. Schaller\altaffilmark{1}, H.G. Roe
\altaffilmark{1}, D.L. Rabinowitz\altaffilmark{2}, C.A. Trujillo\altaffilmark{3}}
\altaffiltext{1}{Division of Geological and Planetary Sciences, California Institute
of Technology, Pasadena, CA 91125}
\altaffiltext{2}{Department of Physics, Yale University, New Haven, CT 06520}
\altaffiltext{3}{Gemini Observatory, 670 North A'ohoku Place, Hilo, HI 96720}
\email{mbrown@caltech.edu}

\begin{abstract}
We have used the Hubble Space Telescope to directly measure the angular size 
of the large Kuiper belt object 2003 UB313. By carefully calibrating the 
point spread function of a nearby field star, we measure the size
of 2003 UB313
to be 34.3$\pm$1.4 milliarcseconds, corresponding to a diameter of 
2400$\pm$100 km or a size $\sim5$\% larger than Pluto. 
The V band geometric albedo of 2003 UB313 is $86\pm7$\%. The extremely high
albedo is consistent with the frosty methane spectrum, the lack of red coloring,
and the lack of observed photometric variation on the surface of 2003 UB313.
Methane photolysis should quickly darken the surface of 2003 UB313, but
continuous evaporation and redeposition of surface ices appears capable
of maintaining the extreme alebdo of this body.

\end{abstract}


\section{Introduction}
The planetary-sized scattered Kuiper belt object 2003 UB313 was discovered 
in an ongoing survey at the Samuel Oschin telescope at Palomar Observatory
\citep{2005ApJ...635L..97B}.  The heliocentric distance of 97.5 AU and V magnitude of 
18.8 at the time of discovery implied a size larger than Pluto for any albedo
lower than 96\%, but the initial discovery was incapable of providing a 
measurement of the albedo. The infrared reflectance
spectrum of 2003 UB313 is dominated
by absorption due to frozen methane \citep{2005ApJ...635L..97B}, like the spectra of Pluto and Triton,
suggesting the likelihood of a high albedo for 2003 UB313 like the
50 - 80\% albedos of these two objects. With such a high albedo
2003 UB313 would have a diameter in the range of 2500 to 3200 km.

To date, most Kuiper belt objects with measured sizes have had their diameters determined through radiometry.  In this technique, the measured thermal infrared flux (or fluxes)
from an object is converted to a diameter using models in which the thermal emission properties and surface temperature distributions of the object are assumed
\citep{2004A&A...415..771A,2005ApJ...624L..53C}. Owing to the uncertainties in the appropriate
model parameters, errors in these size determinations are large.
The object (50000)
Quaoar had its size directly measured by Hubble Space Telescope (HST)
observations \citep{2004AJ....127.2413B} which are unaffected by uncertainties in
thermal emission modeling. The major source of uncertainties in these
measurements is the unknown center-to-limb photometric function of 
Quaoar, which changes the apparent angular diameter of the disk.

Initial Spitzer Space Telescope observations of 
2003 UB313 resulted in a 70 $\mu$m
upper limit of 2 mJy, consistent with a size upper 
limit of 3310 km and lower limit to the albedo of 44\% (Brown et al, in prep).
More recent observations from the IRAM telescope succeeded in detecting
a flux of $1.27 \pm 0.29$ mJy  at 1.2mm, suggesting a size between
2600 and 3400 km and an albedo between 42 and 72\% \citep{Bertoldi}. 
An object of this size would have an angular diameter seen from the earth
of between 37 and 48 mas, similar to or larger than Quaoar,
thus
should be directly resolvable from HST observations. We present here
such observations and the size and albedo of 2003 UB313 directly 
measured by HST.

\section{Observations}
We obtained 16 119-second exposures and
12 133-second exposures of 2003 UB313 using the High Resolution Camera
(HRC) of the Advanced Camera for Surveys (ACS) in the F435W
filter in two orbits of the Hubble Space Telescope (HST). 
The first observations began at 8:29 on 9 December 2005 (UT) and one observation
was obtained every 170 seconds. The second observations 
began at 6:52 on 10 December 2005
 and were obtained every 184 seconds. 
The timing of the observations was chosen
so that 2003 UB313 would be a few arcseconds from a moderately 
bright field star located at 01h35m37.63 -05d38m57.8, and
the telescope was tracked at the sidereal rate so that
accurate point spread function (PSF)
measurements could be obtained from the untrailed image of the star. 
The motion of 2003 UB313 caused
a smearing of about 30 mas of the image of 2003 UB313
during each observation, a
factor which needed to be taken carefully into account in the analysis of
the angular size of the object.

The star had a brightness of $58.8\pm0.2$  
counts s$^{-1}$ ($B=20.152\pm 0.004$) 
in each of the images
and was single, well exposed, and not saturated. 2003 UB313 had a brightness
of $149.9\pm0.7$ counts s$^{-1}$ ($B=19.136\pm 0.005$) and traveled across a 
field uncontaminated by
detectable background
determined from our full PSF fitting analysis,
described below.

The method used to determine the size of 2003 UB313 from these observations
is almost identical to that described and validated
in detail in \citet{2004AJ....127.2413B}. Here, we very briefly
summarize the procedure, and particularly discuss a few small
 improvements.
First, for each of the 28 individual stellar
images, the PSF is modeled using the HST PSF
simulator package TinyTim (Krist \& Hook, 2001) . While TinyTim's
default parameters provide a
close match to the average HRC PSF, the HST PSF is not absolutely stable,
so a fit is performed to the field star in each image.
To obtain a best-fit PSF, we use a downhill simplex $\chi^2$
minimization method 
to fit the first 8 Zernike
terms in the aberration of the TinyTim model at the location of the star.
Figure 1 shows the best-fit focus for the individual images, while Figure
2 gives an example of the best-fit model for one of the images.
In Brown \& Trujillo and in subsequent experiments on other images
we found that the focus varied smoothly over
the course of an orbit. Here we see the same trend but with a few significant
outliers. While some of this scatter is likely real and 
caused by the increase in
jitter with HST now in two-gyro mode, we chose to discard as 
unreliable the images in which the measured focus appears to vary greatly
between adjacent observations (though we note that the final results are
not sensitive to the removal of these points).

With the aberrations measured for each stellar image, TinyTim is used 
to construct a PSF at the location of 2003 UB313, accounting for the 
(small) PSF 
changes due to the field-dependent aberration of the HRC. 
This PSF is convolved 
with the measured motion vector of the object and then convolved with 
model disks with diameters between 10 and 50 milliarcseconds. 
This model image is then convolved with the three-by-three pixel gaussian
kernel representing the effects of CCD charge diffusion.
For each
image a best-fit model disk diameter is found which minimizes the 
sum of the square of the residuals between the modeled image and the data.
An example of a best-fit model is shown in Figure 2.

In this process, the two steps which are the largest source of potential error
are the choice of the charge diffusion kernel and the choice of the 
center-to-limb brightness profile of the model disks. TinyTim provides an 
estimate of the position-dependent charge diffusion kernel on the HRC CCD, 
but this estimate is based on extrapolations from measurements at other
wavelengths \citep{2004SPIE.5499..328K}. 
Because this parameter is critical, we chose to 
independently measure the charge diffusion at the location of our star
and of 2003 UB313. We used archival 
images of the globular cluster M92 obtained with HRC in the
F435W filter and found several
isolated stars near the positions of the central PSF star and the positions
of 2003 UB313 on the two separate orbits.
We used the procedure described above to fit the PSFs of the 
globular cluster stars near the location of the central PSF star, 
and then we transfered this PSF to the locations of 2003 UB313
using TinyTim.
We then performed a $\chi^2$ minimization of the charge diffusion kernel
of the globular cluster stars at the locations of 2003 UB313. This procedure
does not give us a reliable measure of the absolute width of the charge
diffusion kernel (as we have to assume an initial charge diffusion 
kernel for the 
central region), but only a measure of the difference between the center 
location
of the PSF star and 2003 UB313. Fortunately, our measurements are only
sensitive to relative difference and are not dependent on the absolute 
values. We find that the Gaussian
width of the charge diffusion kernel at the positions
of 2003 UB313 are identical for the two measured positions
and are 
$1.04\pm0.01$ times
larger than the width at the center 
PSF star.

The other important factor to consider 
is the center-to-limb brightness profile of the model disks. An object
with a flat center-to-limb profile will appear larger than an object
whose brightness drops steeply near the limb.
In \citet{2004AJ....127.2413B} we considered an extremely wide range of possible profiles
as nothing was known about the surface of Quaoar at the time. For 2003 UB313,
however, we know that the surface is covered in methane frost \citet{2005ApJ...635L..97B} like
the surfaces of Pluto and Triton. As we have precise measurements of
the center-to-limb profile of Triton from well resolved Voyager measurements,
we take this profile, parameterized with a Hapke model \citep{1994Icar..109..296H}, as our 
best analog for the profile of 2003 UB313. 

The size measured for 2003 UB313 in each of the 
17 images not discarded due to 
focus deviations is shown in Figure 1.
Taking into account the random errors and charge diffusion
width uncertainty of 1.0 mas each
we derive an angular size for 2003 
UB313 of $34.3\pm1.4$ mas. At a geocentric distance of 96.4 AU, this angular
diameter corresponds to a diameter of 
$2400 \pm 100$ km. For an absolute
$V$ magnitude of $H_v=-1.12\pm 0.01$ 
the albedo of of 2003 UB313 is $0.85\pm0.07$. 

\section{Discussion}
Though 2003 UB313 is 20\% brighter than Pluto in absolute terms, it is only
6\% larger, as the albedo of 2003 UB313 is even higher than that of Pluto.
This extremely high derived albedo gives us confidence in our choice
of Triton as an analog for the center-to-limb function. Lower albedo
icy satellites have center-to-limb profiles which are much flatter than
that of Triton. Use of an icy-satellite like profile would give a smaller
size and thus even 
higher albedo for 2003 UB313, inconsistent with the low albedos
of the icy satellites. We can, in fact, self-consistently solve for
the center-to-limb profile and the albedo by assuming a Hapke model to
give both
the profile and the geometric albedo. The profile gives a unique measurement
of the angular size, and thus the albedo, but the Hapke model independently
gives an albedo. The only Hapke models which give consistent albedos have 
parameters very similar to Triton, with high single-scattering albedo strongly
backscattering particles and smooth surfaces.

The measured size is well below the Spitzer upper limit of 3310 km, but
apparently 
inconsistent with the reported IRAM measured size of 3000$\pm$400 km
\citep{Bertoldi}.
A reevaluation of the IRAM measurement, however, suggests that the
measurements are indeed compatible. The IRAM size measurement
assumed an absolute magnitude for 2003 UB313 of $H_v=-1.16\pm0.1$, which is 
about 4\% brighter than the best current measurement of $H_v=-1.12\pm0.01$
(Rabinowitz et al., in prep), suggesting a smaller size. 
Reevaluating a thermal model with the fainter absolute magnitude
and assuming that 2003 UB313
is pole-on and is the largest size allowed by our uncertainties predicts
a 1.2mm emission of 0.94 mJy, consistent at the 1.1$\sigma$ level 
with the $1.27\pm0.29$ mJy reported
by IRAM. Even an equator-on orientation would give a flux of 0.70 mJy,
consistent at the 2$\sigma$ level.

The $86\pm7$\% albedo of 2003 UB313 is significantly 
higher than the $\sim$60\% average of Pluto, but 
consistent with the albedo of 
the individual brightest areas on Pluto \citep{2001AJ....121..552Y}.
2003 UB313 is also less red than Pluto \citep{2005ApJ...635L..97B}, and Pluto's darkest
regions appear to be a major source of 
the red color \citep{1999AJ....117.1063Y}. These two 
characteristics suggest that the surface of 2003 UB313 might resemble
the brightest regions on Pluto while containing few if 
any of the darkest regions,
a suggestion consistent with the lack of detectable photometric 
variation on 2003 UB313 (Roe et al., in prep.). 

The geometric albedo of 2003 UB313 is higher than that of all known bodies
in the solar system with the exception of Saturn's satellite Enceladus,
which has a geometric albedo of 103\%
and has active water plumes capable of providing continuous frost
resurfacing (Hansen et al. 2006).
Even 
Triton, which has active geysers and appears
freshly resurfaced, has a slightly lower albedo of 0.77 \citep{1994Icar..109..296H}. Typical
inactive 
water-ice covered satellites in the outer solar system have albedos
ranging from 0.2 to 0.4, with exceptions only related to activity 
or exogenic processes.
On 2003 UB313 even highly reflective methane frosts will irreversibly darken 
due to long-term photolysis;
some uncommon process which continues to maintain a
high albedo on 2003 UB313 is required.

One process unique on large bodies in the Kuiper belt with eccentric orbits
is atmospheric
freeze-out. At 2003 UB313's 38 AU perihelion equilibrium temperature of 
$\sim$43 K,
the vapor pressure over a solid nitrogen surface would be $\sim$30 Pa, 
while
the pressure over a solid methane surface would be $
\sim$0.2 Pa (Lodders and Fegley, 1998). 
At the current distance of
97 AU and equilibrium temperature of 27 K these pressures drop by 6 and 8 
orders of magnitude,
respectively. Any nitrogen or methane atmosphere that exists at perihelion
is essentially completely frozen to the surface by aphelion. 

This 560-year cycle of 
evaporation and then freeze-out of the atmosphere of 2003 UB313
will leave fresh undarkened ices on the top layer at aphelion. 
A 30 Pa nitrogen atmosphere
at perihelion corresponds to a layer of solid nitrogen several centimeters
thick at aphelion, while a 0.2 Pa methane atmosphere corresponds to a 
layer tens of microns thick. These layers would sit on top of any dark
photolyzed methane and would maintain a high albedo for grain sizes smaller
than the surface thickness, which appears likely. At perihleion the dark areas
would be revealed, leading to a Pluto-like appearance. 2005 FY9, which 
also has a surface covered in methane frost
\citep{2005DPS....37.5211B,
2006A&A...445L..35L}
and is currently at a distance of 52 AU, should have a surface temperature
approximately midway between Pluto and 2003 UB313. The presence of 
small but discernible
photometric variations on 2005 FY9 (Roe et al., in prep.) and the evidence
for a small surface coverage of warm dark material (Brown et al., in prep.)
suggests that 2005 FY9 could have undergone partial atmospheric freeze-out,
but that at the temperature of 2005 FY9 the process is not complete as on
2003 UB313. The
process of atmospheric freeze-out, which has long been debated for
Pluto (i.e Stern \& Trafton 1984, Stansberry \& Yelle 1999, Eliot et al. 2004)
can now be studied in a growing population of methane rich objects at
a wide range of distances and temperatures in the outer solar system.

{\it Acknowledgments:} We would like to thank the director and staff 
at STScI for providing the opportunity to make these measurements 
and the assistance in making them
happen. This research has been supported by a
grant from STScI and from NASA Planetary Astronomy.

\clearpage

\begin{figure}
\plotone{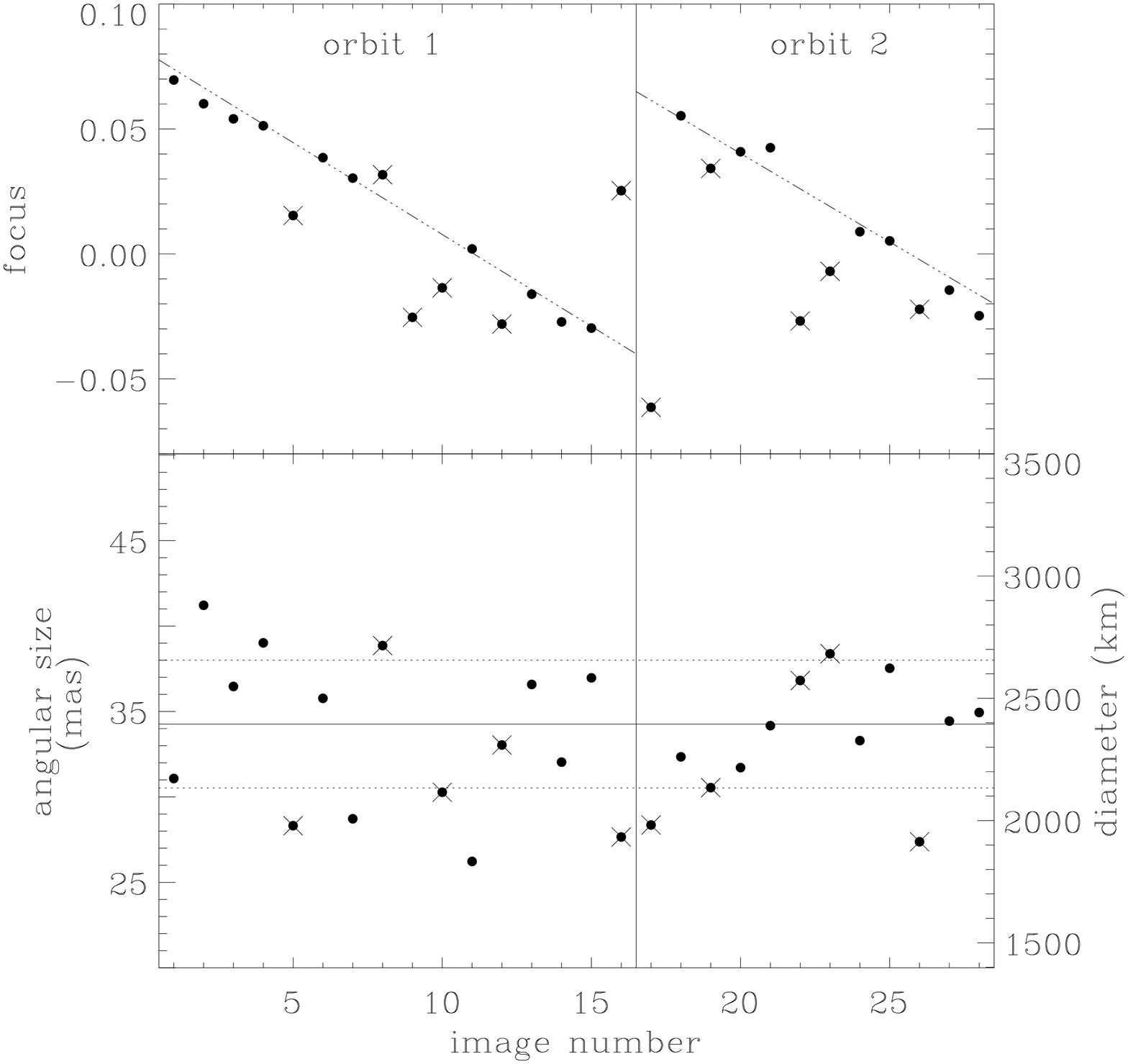}
\caption{The top panel shows the best-fit focus (in units of RMS waves
of aberration at 547 nm) for the 28 images. Previous experience has
shown that the focus varies smoothly over the course of an orbit. We 
expect that the large jumps in derived focus are a product of the 
lower signal-to-noise of the PSF star and of the increased jitter in
two gyro mode. We discard as potentially unreliable all of the measurements
marked with a ''X'' in which the focus appears to shift significantly
between images. The bottom panel shows the best fit disk size for each of the
individual images. The mean is shown as the solid line and the standard 
deviation is shown as the dashed line.}
\end{figure} 

\clearpage

\begin{figure}
\plotone{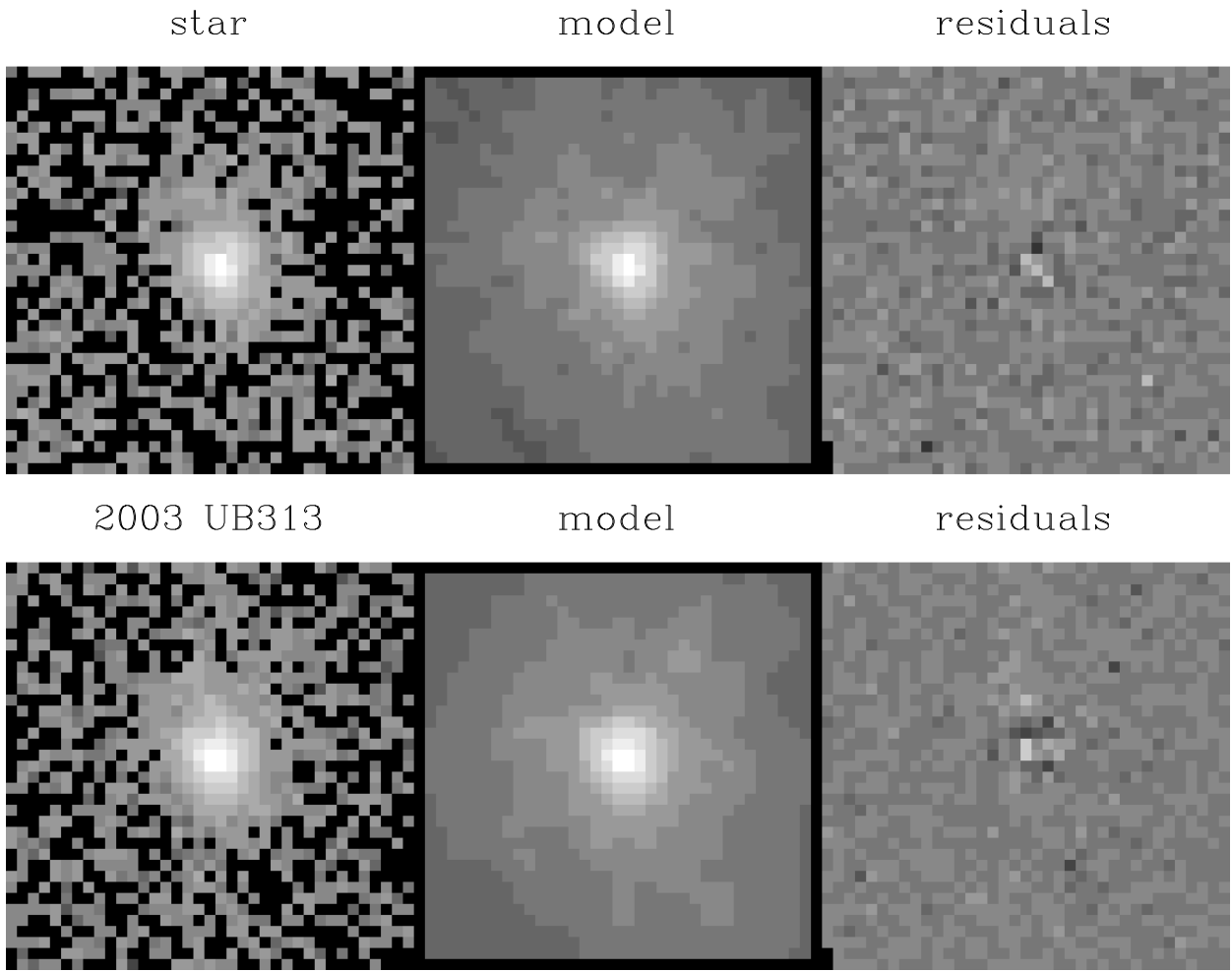}
\caption{Comparison of the star and 2003 UB313 in image 21 with 
the best-fit models. The data and models share identical 
logarithmic scales. The residuals are linearly scaled from 
$\pm5$\% of the maximum of the model image.}
\end{figure}

\end{document}